# Semi-permeable vesicles composed of natural clay

*Anand Bala Subramaniam[1], Jiandi Wan[2], Arvind Gopinath[3] and Howard A. Stone[2]*

1. School of Engineering and Applied Sciences, Harvard University. Cambridge, MA 02138 2. Department of Mechanical and Aerospace Engineering, Princeton University, Princeton, NJ 08544. 3. Fischer School of Physics, Brandeis University, Waltham MA 02453.

## Abstract

We report a simple route to form robust, inorganic, semi-permeable compartments composed of montmorillonite, a natural plate-like clay mineral that occurs widely in the environment. Mechanical forces due to shear in a narrow gap assemble clay nanoplates from an aqueous suspension onto air bubbles. Translucent vesicles suspended in a single-phase liquid are produced when the clay-covered air bubbles are exposed to a variety of water-miscible organic liquids. These vesicles of clay are mechanically robust and are stable in water and other liquids. The formation of clay vesicles can be described by a physical mechanism that recognizes changes in the wetting characteristics of clay-covered air bubbles in organic liquids. The clay vesicles are covered with small pores and so intrinsically exhibit size-selective permeability, which allows spontaneous compartmentalization of self-assembling molecules in aqueous environments. The results we report here expand our understanding of potential paths to micro-compartmentalization in natural settings and are of relevance to theories of colloidal aggregation, mineral cycles, and the origins of life.

# Introduction

Biological cells are compartmentalized by semi-permeable phospholipid membranes that separate an aqueous interior from an aqueous exterior. How Nature arrived at compartments composed of complex phospholipids as the ubiquitous container of living cells, and if life began from self-organizing molecules compartmentalized in simpler structures formed in the environment, are some of the fundamental questions for the scientific rationalization of the origins of life [1-10]. Meanwhile, many synthetic strategies have been developed to manufacture semi-permeable compartments for technological applications: compartments have been synthesized from alternating layers of polyelectrolytes (polyelectrolyte capsules) [11], block co-polymers (polymersomes) [12], and polymeric colloidal particles (colloidosomes) [13-15].

The collective knowledge gained from designing these well-known artificial compartments is however of limited use in elucidating possible routes to compartments in natural settings. For example, (i) polyelectrolyte capsules are made by templating synthetic polyelectrolytes on synthetic polymeric particles [11], (ii) polymersomes are manufactured with synthetic block co-polymers [12], and (iii) the many examples of colloidosomes and their variants are manufactured with synthetic polymeric particles or surface-modified particles, and require binding, annealing and/or centrifugation steps [13-16]. Clearly, all of these compartments require man-made components and employ complex steps as an integral aspect of the fabrication pathway, which, while useful for tailoring properties such as permeability and size, exclude the possibility that such compartments could arise in the environment.

In this paper we report that robust thin-shell semi-permeable compartments can be formed through a relatively simple route from the clay montmorillonite. We focus on montmorillonite among the many naturally occurring colloidal minerals since montmorillonite is one of the more important clay minerals, present widely on the Earth [17], and studied in fields as diverse as colloid [18] and materials science [19, 20], earth [21, 22] and exoplanetary science [23, 24], mineralogy [17], medicine [25, 26], catalytic chemistry [27, 28], and, more importantly, has been shown to play a range of roles in potential origins of life scenarios. For example, montmorillonite is known to catalyze the vesiculation of fatty acids [3] and oligomerization of activated nucleotides [29], and has been shown to protect polynucleotides from ultra-violet radiation damage [30]. While montmorillonite is a specific species of mineral grouped under the general category of clay minerals (illite, kaolinite, and attapulgite are other examples of minerals classified as clays) [17], in this paper we use the word 'clay' to refer specifically to montmorillonite.

We find that (i) thin layers of unmodified natural montmorillonite can be assembled onto gas bubbles in water through mechanical forces, forming armored bubbles [31] of clay, (ii) when the clay armored bubbles are exposed to a variety of water-miscible organic liquids, a nanoscale wetting transition, confirmed thorough bulk wettability studies, triggers the formation of *clay vesicles*, which are mechanically robust multi-layer yet thin-shell aggregates of montmorillonite suspended in a permeating single-phase liquid, and (iii) we confirm that the clay vesicles demonstrate size-selective permeability and show that they support spontaneous self-organization and compartmentalization in the presence of self-assembling molecules. These vesicles of clay are a class of compartments distinct from previous reports of inorganic compartments, which are generally tubular precipitates of metal salts [7, 32].

While montmorillonite has been widely studied for catalytic and chemical roles, our report that montmorillonite can form semi-permeable compartments is the first indication that this clay could have also played a *structural* role in the origins of life. Indeed, more generally, this is the first demonstration (to the best of our knowledge) that a natural colloidal mineral can organize into structured inorganic compartments through a robust and simple pathway, which may conceivably be useful in other materials science applications.

## Results and Discussion

**Formation and characterization of armored bubbles from unmodified natural montmorillonite**

Bulk emulsification methods are reported to be successful for producing particle-stabilized emulsions and foams with hydrophobic, synthetic or surface-modified clays[33-35]. We do not observe bubbles covered with a layer of montmorillonite when a dilute (~ 1 percent wt/wt) aqueous suspension of clay (particle width < 2 micron, thickness ~ 1.5 nanometers [17, 18, 36]) is emulsified by hand or with a vortex mixer. This result agrees with previous observations of interactions between unmodified montmorillonite and air/water interfaces [37]. Clay armored bubbles are produced (approximately 10-100 bubbles per attempt) when a suspension of clay containing bare air bubbles is sheared manually between two clean glass slides (Fig. 1A-C). Clay armored bubbles can also be produced when bubbly clay suspensions are sheared between other hydrophilic surfaces such as aluminum, copper, stainless steel, and quartz. This observation suggests that the side-wall material may be of secondary importance to the hydrodynamic conditions in the gap between the two surfaces for the formation of armored bubbles from clay. Note that since the process is manual we do not control the distance between the plates (i.e. the gap width) in our experiments. We also emphasize that the montmorillonite particles are not modified chemically in any way.

While the hydrodynamics of the multiphase fluid suspension in the gap between the surfaces is a complex combination of shear and extensional flows, we envision that the motion of the slides likely orients the anisotropic plate-like particles parallel to the boundaries and promotes capture of the particles at the wedge between the air-water interface and the rigid surface (Fig. 1D). The narrow gap also deforms the air bubble, which increases the interfacial area available for particle attachment. Once adsorbed on the interface, the particles are trapped in a deep potential well [38], which eventually enables full coverage of the bubble surface (Fig. 1E). It is apparent that this method, which provides targeted delivery of particles and a larger area of interaction, improves the otherwise inefficient process of clay nanoplate capture and makes the formation of clay armored bubbles observable within the laboratory timeframe.

Scanning electron microscopy (SEM) imaging of air-dried (Fig. 1F) clay armored bubbles reveal that the clay layer appears smooth and continuous with no observable holes (Fig. 1G). The wall is composed of multilayers of overlapping clay nanoplates with a thickness that ranges from 5 to 10 nm as measured by high resolution transmission electron microscopy (HRTEM) (Fig. 1H). The stability of the clay shell to drying indicates that the clay layer is self-supporting and does not require capillary forces to maintain structure. Pickering emulsions [13, 14] and armored bubbles [31] produced with spherical particles generally disintegrate into individual particles when dried, i.e. when no fluid/fluid interface is present holding the particles together. In fact, clay armored bubbles can survive multiple drying and rehydration cycles without apparently losing integrity (Fig. 2).

Can the observed resilience of the clay armored bubbles to drying-induced disintegration inform us of the interactions between the clay nanoplates in the shell? In aqueous suspension of low ionic strength (matching the conditions in our suspensions), montmorillonite separates into individual nanoplates that are colloidally stable due to electrostatic repulsions [18, 36]. We suggest that the mechanical conditions in the gap during the process of armored bubble formation provides sufficient energy to overcome repulsions and allows the particles to access their primary van der Waals minimum, causing the permanent aggregation of the clay nanoplates on the bubble surface. The aggregation of the particles is what allows the shell to be self-supporting (i.e. the shell does not collapse under its own weight) in the absence of a liquid/gas interface.

Simple calculations show the feasibility of this scenario. The weight of a 6 nm thick clay shell (density of montmorillonite 2010 kg/m$^3$) on an armored bubble 25 μm in radius is $O(10^{-12})$ N. The van der Waals adhesion force [39] between two nanoplates is $F_{vdW} = \frac{A}{6\pi D^3} \cdot area$, where $A$ is the Hamaker constant = 7.3 x 10$^{-21}$ J for montmorillonite [22, 36] and $D$ is the interparticle distance. Taking the area of the plates to be 1 μm$^2$, and for $D$ = 0.2 nm, the typical separation distance for solids in adhesive contact [39], $F_{vdW}$ is $O(10^{-5})$ N. For $D$ = 1.5 nm (typical thickness of a nanoplate), $F_{vdW}$ is $(10^{-7})$ N, which is still much larger than the weight of the shell. The van der Waals adhesion forces calculated here are lower bounds for the cohesive interactions since the shell is actually composed of multiple interacting plates in an overlapping structure. Thus not only is aggregation sufficient to support the shell in the absence of capillary forces, clay aggregation can also explain the resilience of the shell to multiple drying and rehydration cycles. The aggregation of the nanoplates on the clay armored bubbles is an important observation that we will return to further below.

**Thin-shell vesicles of clay form when clay armored bubbles are exposed to certain water-miscible organic liquids**

When clay armored bubbles are exposed to the common water-miscible organic liquid ethanol (continuous phase: 99 percent ethanol, 1 percent water and armored bubbles), the inner gas phase dissolves leaving behind the montmorillonite shell as a stable translucent vesicle (Fig. 3A-C). The vesicles can be transferred back into pure water without losing integrity (Fig 3D). Vesicles sealed in glass capillaries remain unchanged in ethanol or water for at least a year. The vesicles are also stable in aqueous solutions at a wide range of ionic strengths and pH (Experimental details). Localized imposed deformations with glass capillaries reveal that the clay vesicles are mechanically robust and respond plastically without disintegration (Mov. S1, S2), while atomic force microscope (AFM) measurements indicate a bending rigidity $\kappa$ that ranges from $4 \times 10^{-16}$ J to $11 \times 10^{-16}$ J. For comparison, gel-phase phospholipid vesicles (lipid vesicles are composed of bilayers of lipid molecules ~ 6 nm thick, and the lipids crystallize due to relatively weak intermolecular van der Waals forces [39] ) have $\kappa = 1 \times 10^{-18}$ J [40].

Clay vesicles can also be formed through an alternate pathway that does not require the bubbles to be surrounded by water. As we have shown above, clay armored bubbles that are dried and rehydrated with water trap air (Fig. 2). However, when the dried clay aggregates are rehydrated with ethanol, internal air bubbles dissolve to produce clay vesicles that remain adhered to the surface (Fig. 3E-H).

Naturally, we wondered if clay vesicles formed in other types of water-miscible organic liquids. We conducted experiments with representative organic compounds, a carboxylic acid - acetic acid $CH_3COOH$, a ketone – acetone $(CH_3)_2CO$, an aldehyde – acetaldehyde $CH_3CHO$, an amide – formamide $HCONH_2$, a sulfoxide – dimethylsulfoxide $(CH_3)_2SO$, a furan – tetrahydrofuran $C_4H_8O$, heterocycles – morpholine $C_4H_9NO$ and pyridine $C_5H_5N$, a triol- glycerol $C_3H_5(OH)_3$, a glycol - ethylene glycol $C_2H_4(OH)_2$ and alcohols – methanol $CH_3OH$ and ethanol $C_2H_5OH$. To standardize concentrations across the experiments, we use 10 μl of an aqueous suspension of armored bubbles and 990 μl of the respective organic liquids (i.e. the continuous phase is 99 percent organic liquid). Clay vesicles do not form when clay armored bubbles are exposed to glycerol, formamide or for bubbles remaining in pure water, while clay vesicles are formed when the bubbles are exposed to acetic acid, acetone, acetaldehyde, methanol, ethanol, ethylene glycol, pyridine, morpholine, dimethylsulfoxide and tetrahydrofuran. In all cases, the vesicles once formed, can be transferred into pure water and into all the other water-miscible organic liquids tested above.

**Characterization of clay vesicle structure**

Under the SEM the vesicles generally appear as spheroidal wrinkled shells perforated with irregularly shaped submicron pores, as opposed to the smooth pore-free structure of clay armored bubbles (Fig. 4A, Fig. 4B, compare with Fig. 1F,G). The structure of individual clay nanoplates is shown in Fig. 4C. The thin-walled hollow structure of the vesicles can be seen in Fig. 4D and a close up of the torn edge shows that the walls are not more than 10 nm thick (Fig. 4E). While exposure to ethanol leads to the formation of pores on the surface of the vesicles, HRTEM imaging reveals that there is no observable reorganization of the aggregated clay nanoplates within the shell (Fig. 4F). Similar to dried clay armored bubbles, the walls of the clay vesicles are also composed of between four to six parallel overlapping clay nanoplates, with a wall thickness that ranges from 4 to 8 nm.

**Proposed mechanism of formation of clay vesicles**

Why do clay vesicles form when clay armored bubbles are exposed to certain organic liquids? The formation of clay vesicles is general for different types of organic liquids and does not correlate with specific chemical functional groups of the organic compounds. Thus, a physical rather than chemical mechanism is likely to be dominant.

Next, we recognize that the solubility of air in polar organic liquids is about 10 times greater than the solubility of air in water [41]. Naively this might explain our observation of the rapid dissolution of the gas phase of clay armored bubbles and the subsequent formation of clay vesicles when bubbles are exposed to organic liquids. We thus seek to test the hypothesis that increased gas solubility can explain the formation of clay vesicles. Clay armored bubbles were prepared normally, i.e. in water that is saturated with air. 10 µl of this suspension was then added to 990 µl of water that was degassed in a vacuum chamber. The clay armored bubbles dissolve rapidly and exhibit nonspherical shapes as dissolution proceeds. We never observe clay vesicles forming. Instead aggregates of clay are ejected from the bubble surfaces and compact clumps of clay are left behind after all the air has dissolved (Fig. 5). Thus, we conclude that a change in the rate of gas transfer is not the primary mechanism for the formation of clay vesicles.

Having excluded the above potential explanations, the formation of clay vesicles can be rationalized by a transition from partial to complete wetting of the clay aggregate by the outer liquid. Our proposed mechanism of formation is illustrated schematically in Figure 6. We begin by first considering the particles on clay armored bubbles which are mostly permanently aggregated in an overlapping multilayered structure. We posit that while most of the nanoplates on the clay armored bubbles are indeed aggregated in their van der Waals minimum, a population of nanoplates in 'defect sites' are outside of their minimum interaction potential (the red particles in the schematic) (Fig. 6A). This supposition is reasonable since the process of armored bubble formation and clay aggregation is uncontrolled, and physical and chemical heterogeneities [17, 18] should cause differences in interaction energy between the nanoplates.

Next, we extend the concept of the spreading parameter, $S = \gamma_{SV} - (\gamma_{LV} + \gamma_{SL})$, where $\gamma_{SV}$, $\gamma_{LV}$, and $\gamma_{SL}$ are the solid-vapor, liquid-vapor, and solid-liquid surface energies per unit areas respectively, which describes the thermodynamic criterion for equilibrium wetting of chemically homogenous smooth solids [39, 42]. On bulk solids, when $S \geq 0$, the liquid displaces the vapor phase and wets the solid completely, forming a wetting film on the solid. When $S < 0$ only partial wetting occurs and the liquid forms an equilibrium drop on the solid. We expect a similar response in our system, for vesicle-forming liquids $S \geq 0$, while for those liquids that do not produce vesicles $S < 0$.

In partially wetting liquids, $S < 0$ (i.e. in water, formamide, glycerol), the outer liquid does not penetrate the aggregated clay layer and the liquid forms a 'contact angle' on the solid (Fig. 6A). The enclosed air phase is thus stable and the defect particles are unaffected. This view is supported by the observation that the clay shell remains free of pores even after multiple drying and rehydration cycles with water. When $S \geq 0$ which occurs when the clay armored bubbles are exposed to completely wetting liquids, the outer liquid spreads and displaces weakly adhered nanoplates (Fig. 6B). The voids created by the displaced plates at the putative defect sites accounts for the irregularly shaped submicron pores observed on the vesicle surface (i.e. Fig. 4A, B). Once the outer liquid spreads completely and displaces the air from the inner surface of the clay aggregate (Fig. 6C), the air bubble is no longer stable [15] and dissolves into the surrounding liquid leaving behind a clay vesicle (Fig. 6D).

The observation that the majority of the particles in the clay layer remain unaffected by the wetting transition is consistent with well-known observations that colloidal particles aggregated irreversibly in their primary van der Waals minimum are difficult to disrupt without significant mechanical agitation [39, 42]. Although natural montmorillonite is chemically and physically heterogeneous, the generality of vesicle formation for liquids of distinct chemical and physical properties suggests that effects such as contact angle heterogeneity or pinning are unlikely to be the primary determinant in clay vesicle formation.

**Test of the wetting hypothesis**

Our hypothesis predicts that vesicle-forming liquids should have spreading parameter, $S \geq 0$ and spread more on the clay, while for those liquids that do not produce vesicles $S < 0$ and should spread less on the clay. We employ three methods to characterize the wetting properties of montmorillonite, (i) contact angle measurements of the liquids on compressed clay tablets, (ii) measurements of the maximum extent of spreading by the liquids on the surface of compressed clay tablets, and (iii) calculations of $S$ from solid, liquid and vapor surface energy components.

$S$ can be defined in terms of the equilibrium contact angle, $\theta_e$ and $\gamma_{LV}$, to give the well-known Young-Dupre equation $S = \gamma_{SL}(1 - \cos\theta_e)$ [39, 42]. Liquids that spread completely have $\theta_e = 0$, and $S \geq 0$ (however the Young-Dupre equation does not admit positive solutions for $S$). We probed wetting behavior by depositing a small volume of the respective liquids on compressed montmorillonite tablets. Some liquids form drops (Fig. 7A-C) while others spread extensively on the surface (Fig. 7D, E). All liquids also imbibe into the clay tablet and evaporate, and eventually (within minutes) appear to 'dry' off the surface. Apparent contact angles could be measured for water, glycerol and formamide (Fig. 7A-C). All other liquids appear to spread with very small dynamic angles (< 3°, experimental resolution) indicating a zero equilibrium contact angle (Fig. 7D, E). Based on the observed angles, we calculate an effective $S$ (Fig. 7F). Consistent with our hypothesis, the three liquids that do not form vesicles - water, formamide and glycerol - exhibit partial wetting on the clay tablet, while all of the other liquids exhibit complete wetting on the clay tablet.

After the liquid has 'dried' off of the surface, distinct regions of different optical intensity are left behind on areas that were in contact with liquid (Fig. 7A-E right column). We use this observation to further quantify wetting behavior by measuring the maximum extent spread by the liquids on the surface of the clay tablet and plot the results in ascending order in Figure 7G. The three liquids that do not from vesicles spread the least, while liquids the form vesicles spread more extensively, with morpholine and acetic acid spreading over the entire surface of the tablet (Experimental Details, Fig. 10).

Finally, we calculate $S$ by using surface energy components of the three phases and the surface energy definition, $S = \gamma_{SV} - (\gamma_{LV} + \gamma_{SL})$. The surface energy of the liquids, $\gamma_{LV}$, was measured using the pendant drop method (Table 1). Literature values [36, 43] for the surface energy of montmorillonite with an adsorbed layer of water (i.e. montmorillonite exposed to atmospheric air, which matches our experiments), $\gamma_{SV}$, range from 44.2 mJ/m$^2$ to 49.5 mJ/m$^2$. The third component of the spreading parameter was obtained using an equation of state [44] that relates $\gamma_{SV}$ and $\gamma_{LV}$ of pure liquids (our experiments have a continuous phase that is ~ 99 percent organic liquid) to $\gamma_{SL}$. We plot in ascending order the calculated $S$ for each liquid (Fig. 8). Again consistent with our hypothesis, calculations show that $S > 0$ for vesicle-forming liquids, while for non-vesicle-forming liquids $S < 0$. Note that the calculated magnitudes of the negative $S$ parameters do not match the effective $S$ values obtained through apparent contact angle measurements using the Young-Dupre equation (Fig. 7F), which is not unexpected for a complicated material such as natural montmorillonite. However the *sign* of $S$, which determines if the free-energy change is favorable for complete wetting, is consistent between both methods. A comparison of the magnitude of the positive $S$ values cannot be made due to the limitations of the Young-Dupre equation. Since three methods agree with the wetting criterion, we conclude that a wetting transition is the primary mechanism for clay vesicle formation in organic liquids.

**Confirmation of semi-permeability and demonstration of spontaneous compartmentalization**

Since the vesicles are perforated with submicron pores, we reasoned that they could exhibit size-selective permeability. The selective permeability of the clay vesicles combined with the self-assembling properties of molecules such as polymers, lipids, and surfactants, could lead to the spontaneous emergence of compartmentalized microenvironments; small precursors can diffuse freely into or out of the clay vesicles, while larger condensed structures are trapped when they exceed the size of the pores on the vesicle walls. If allowed to evolve over time, the selective permeability will eventually lead to a higher density of large condensed structures within the spatial confines of the vesicles

To test our hypothesis for vesicle-enclosed self-assembly, we conduct experiments with a minimal system consisting of a single molecular species, the fatty acid salt sodium oleate, which self-assembles into liposomes under alkaline conditions (4). A micellar solution of the fatty acid is added to an aqueous solution, buffered at pH 8.5, containing surface-adhered clay vesicles (Fig. 9A). The fluorescent polymer rhodamine-dextran (hydrodynamic diameter = O(10) nm) is used to label the liposomes, while 110 nm diameter fluorescent polystyrene particles are used as size standards. The rhodamine-dextran and fluorescent particles are visualized simultaneously using multi-channel confocal fluorescence microscopy. After an incubation time of about 20 minutes, micrometer-sized oleate liposomes are observed everywhere in solution including in the interior of the clay vesicles (Fig. 9B). In the green channel, the interior of the clay vesicle appears dark against a bright background of the 110 nm particles (Fig. 9C).

Strip intensity profiles through the respective images clearly demonstrate the selective permeability. Fig. 9D shows that the intensity of the rhodamine-dextran is the same on the inside and outside of the vesicle, which indicates that the small polymer permeates the clay vesicle, while Fig. 9E shows that the intensity of the larger 110 nm particles is approximately zero in the vesicle interior, which illustrates size exclusion (about 75 percent of the vesicles are impermeable to 110 nm particles in a sample of 115 vesicles, while 2 vesicles were found to be impermeable to fluorescently-labeled dextran with an average Stokes diameter of 17 nm). Note that the intensities of both fluorescent probes on the vesicle wall are more than an order of magnitude greater than the intensities in solution, highlighting the well-known ability of clays to adsorb and concentrate dissolved materials [17].

A superimposed image of the red and green channels further emphasizes that the clay vesicle is impermeable to 110 nm objects and, by implication, larger particles, yet encloses micrometer-sized liposomes (Fig. 9F). We conclude that oleate monomers and micelles must have diffused through the vesicle wall and self-assembled to form the enclosed liposomes, confirming that clay vesicles can indeed compartmentalize condensed structures of molecular precursors. Dynamic populations of liposomes can clearly be seen in the exterior and the interior of the vesicles when the oleate concentration is high (Fig. 9G, Mov. S3). Large liposomes that form in the inside of the vesicles are observed never to leave, while liposomes formed on the outside are barred from entering. Eventually, over the course of several hours, more complex lipid mesophases form in the interior of the vesicles (Fig. 9H, Mov. S4). It is thus clear that the clay vesicles naturally support compartmentalization and evolution of complex internal structures of even this simple self-assembling system.

## Conclusions

Looking forward, further work needs to be done on characterizing and optimizing armored bubble formation, characterizing the detailed mechanism for interfacial attachment, and determining if other natural colloidal minerals can also be assembled into armored bubbles and vesicles. Furthermore, while amphiphilic surfactants, such as Triton X-100, sodium dodecyl sulfate, and myristic acid also form vesicles when added to clay armored bubbles in water (Experimental Details) we leave the more complicated study of the effects of surface-active molecules for future work.

To summarize, our results demonstrate that the natural clay montmorillonite can be assembled into robust naturally semi-permeable vesicles in the laboratory. Mechanical conditions for the formation of clay armored bubbles in flows similar to those investigated here might have occurred in gaps between sliding rocks or between pebbles under the action of tidal waves. Simple surfactants and organic liquids with the appropriate wetting characteristics for forming clay vesicles are believed to have been present on the early Earth [45] and other astronomical bodies [46, 47]. The demonstration that vesicles can form from dried clay aggregates (Fig. 3E-H) suggests that these stable aggregates could persist, for example at the shore-line, until the appropriate conditions for vesicle formation occurred. Alternatively, aerosol droplets containing clay armored bubbles or dry aerosols of clay aggregates might be transported to locations that are conducive to the formation of clay vesicles. Indeed, clay vesicles, or other yet to be discovered types of compartments assembled from natural nanoparticles, may have served as simple primitive inorganic precursors to organic proto-cells.

Finally, in the present age, the abundant sources of organic liquids and surfactants in the environment, due to microbial [48] and human action [49], may make the formation of clay vesicles more likely. In this context, montmorillonite aggregation in the form of clay vesicles could have implications for geological mineral cycling [21, 22] and might provide protected habitats for suitably-sized organisms.

## Experimental Details

**Materials**

SWy-2 Na-rich montmorillonite was obtained from the Source Clays Repository of The Clay Minerals Society. The organic liquids ethanol (anhydrous, purity ≥ 99.5 %), methanol (anhydrous, purity ≥ 99.8 %), acetone (HPLC grade, purity ≥ 99.9 %), morpholine (ACS reagent grade, purity ≥ 99.0 %), pyridine (anhydrous, purity ≥ 99.8 %), ethylene glycol (anhydrous, purity ≥ 99.8 %), acetic acid (ACS reagent grade, purity ≥ 99.7 %), formamide (for molecular biology, purity ≥ 99.5%), glycerol (SigmaUltra grade, purity ≥ 99.0 %), acetaldehyde (ACS reagent grade, purity ≥ 99.5 %), dimethylsulfoxide (ACS reagent grade, purity ≥ 99.9 %), tetrahydrofuran (anhydrous, purity ≥ 99.9 %), were all purchased from Sigma-Aldrich.

**Preparation of clay armored bubbles**

In a typical experiment, 0.1 g of montmorillonite was suspended in 9.9 g of ultrapure water (Millipore). To disperse the nanoparticles, the suspension was sonicated using a probe sonicator (Branson) for 10 minutes at 10 percent power. The polydisperse suspension was allowed to sediment overnight to obtain mostly particles < 2 μm in width. 100 μl of the suspension was placed onto a pre-cleaned borosilicate glass slide. A second glass slide was placed onto the first slide trapping gas pockets in the suspension between the two slides. The slides were moved relative to each other to shear the suspension containing the air bubbles. Then, the slides were pulled apart and a pipette was used to transfer the bubbles now covered with a layer of the clay particles (clay armored bubbles) into a clean Eppendorf tube containing ultrapure water. The transfer process helped reduce the amount of free particles in solution, facilitating optical visualization and subsequent studies. The process above was repeated until the desired amount of armored bubbles was obtained. The process above was also repeated with silicon wafers, copper plates, stainless steel sheets, and between quartz slides. Armored bubbles could be formed in all these cases. Dried clay armored bubbles were produced by simply drying an aqueous suspension of the bubbles onto a substrate.

**Imaging**

*Scanning electron microscopy (SEM):* Images were obtained using a Zeiss Supra55 Field Emission SEM. Clay armored bubbles were adhered to a glass slide with a thin layer of UV curable glue (Norland Optical), transferred into pure ethanol to produce clay vesicles, and then critical point dried. The vesicles were sputter-coated with a thin layer of platinum prior to imaging.

*Cryo-SEM:* A suspension containing armored bubbles was introduced onto a copper sample grid and flash frozen in liquid ethane at -180 °C. The resultant material containing amorphous ice was loaded into the SEM chamber and allowed to sublimate for an hour. This was sufficient to partially expose clay armored bubbles close to the amorphous ice surface. Samples were imaged in a Phillips Leo SEM.

*Transmission electron microscopy (TEM):* Surface-adhered vesicles were produced on a holey-carbon grid and critical point dried. Torn vesicle edges were imaged using a JEOL 2100 TEM.

*Optical and confocal fluorescence microscopy:* Images and movies were obtained using an upright confocal microscope (Zeiss LSM 510) with a 63x/ 1.0 N.A. water dipping objective. Strip intensity profiles of the confocal images were obtained in ImageJ. Raw data were normalized by dividing all the data points with the average intensity of the first 10 data points. The normalized data were plotted in Origin Pro (ver 8, Origin Lab Corp). Optical microscopy was performed with a Leica DM-IRB inverted microscope equipped with long working distance objectives, and images were captured with a high-resolution CCD camera (Retiga 2000R).

**Atomic Force Microscope (AFM) measurements of the mechanical properties**

Elastic properties of critical point dried clay vesicles were characterized by indentation measurements on a MFP-3DCoax AFM coupled with an inverted microscope (Asylum Research, Santa Barbara, CA). A silicon nitride probe (MikroMash, OR) with a stiffness of 0.15 N/m was used. Raw data were converted into force versus deformation curves using software by Asylum Research. The elastic modulus of the clay vesicles was determined using the method reported by Glynos et al.[50] Briefly, straight lines were fitted to the linear regime of the force-deformation curves (less than 6 nm of deformation, i.e. when deformation is smaller than the thickness of the vesicle walls) in Matlab (ver. 7.1, Mathworks) using the linear least squares method. The slope of the fitted line is the effective spring constant of the vesicle, $k_{eff}$. $k_{eff}$ is related to the Young's modulus, E by [50]:

$$k_{eff} = \frac{4E}{\sqrt{3(1-\upsilon^2)}} \frac{h^2}{R} \quad (1)$$

where $\nu$ is the Poisson ratio, $h$ is the thickness of the wall, and R the radius of the vesicle. Since the clay vesicles are thin rigid shells which primarily deform due to bending (rather than stretching) when subject to localized mechanical forces, we calculate the bending rigidity, $\kappa$ of the vesicles. The bending rigidity also allows us to compare the mechanical properties of the clay vesicles with gel-phase lipid vesicles. Combining the well-known thin-shell equation for the bending rigidity [51],

$$\kappa = \frac{Eh^3}{12(1-\upsilon^2)} \quad (2)$$

with equation (1) gives

$$\kappa = \frac{\sqrt{3}k_{eff}hR}{48(1-\upsilon^2)^{1/2}}$$

we take $\nu = 0.5$ for an incompressible solid.

**Contact angle and fractional area spread measurements**

Montmorillonite powder was pressed using a tabletop hydraulic press (Carver, Inc) in a 13 mm diameter die at a nominal pressure of 15,000 PSI. The resulting tablet was solid with a glossy surface. Tablets were used immediately after pressing. 5 µl of the respective liquids were placed on the tablet and time-lapse images were obtained with a high resolution CCD camera (Retiga). Images at 750 milliseconds after deposition of the liquids were used to measure the contact angles in ImageJ. Images of the maximum area spread by the liquids on the surface of the clay tablet were obtained after 24 hours. Experiments were repeated 10 times per liquid for both the contact angle and spread area measurements.

**Spreading parameter calculations from surface energy components**

$\gamma_{LV}$: Surface tension was measured using the pendant drop method. Drop images were analyzed and fitted with a custom routine in Matlab. Our measured values (Table 1) compared favorably to tabulated surface tension values in the CRC Handbook of Chemistry and Physics [52].

$\gamma_{SV}$: Wu reports the surface energy of montmorillonite as 49.4 mJ/m$^2$ [43], while Duran et al. report the energy as ranging from 44.2- 49.5 mJ/m$^2$, the lower value assuming no acid-base contribution to $\gamma_{SV}$ [36]. For our calculations we take $\gamma_{SV}$ = 49.4 mJ/m$^2$.

$\gamma_{SL}$: Solid-liquid surface energies $\gamma_{SL}$ were determined using an equation of state proposed by Neumann et al. [44] which is valid for pure non-adsorbing liquids, and surfaces with surface energies less than that of pure water, i.e. < 72 mJ/m$^2$:

$$\gamma_{SL} = \frac{\left((\gamma_{SV})^{\frac{1}{2}} - (\gamma_{LV})^{\frac{1}{2}}\right)^2}{1 - 0.015(\gamma_{SV}\gamma_{LV})^{\frac{1}{2}}}$$

where $\gamma_{SV}$ is the solid-vapor surface energy and $\gamma_{LV}$ is the liquid-vapor surface energy.

**Compartmentalization of oleate liposomes**

Stock solutions of: 1 mg/ml sodium oleate (purity ≥ 99.0 %, Sigma) in ultrapure water, 10 mg/ml rhodamine-dextran (average MW 4,000, Sigma) or fluorescein-dextran (average MW 150,000, Sigma) in ultrapure water, 110 nm diameter surfactant-free yellow-green fluorescent polystyrene particles (coefficient of variation: 9.6 % , IDC Corporation) at 0.1 volume percent in ultrapure water, were prepared. In a typical experiment, clay vesicles in 400 μl of 0.2 M bicine solution (Sigma) at a pH of 8.5 were placed in a chamber constructed out of a PDMS gasket adhered to a glass slide. 40 μl of the dextran stock solution, 40 μl of the fluorescent particle stock solution, and 40 μl of the sodium oleate stock solution were added to the chamber and gently mixed with a pipette. The sample was allowed to incubate for 20 minutes before visualization. Experiments omitting either the dextran or the fluorescent particles, or with higher concentrations of sodium oleate were also performed. 4 samples, with 110 nm diameter fluorescent particles or fluorescent 150,000 MW dextran in the exterior, containing a total of 115 vesicles were imaged to determine average permeability.

**Stability of vesicles in conditions of varying pH and ionic strengths**

Montmorillonite clay nanoplates have an unequal charge distribution. The faces are negatively-charged, while the edges are positively-charged at neutral and acidic pH (4, 5). We changed the pH of aqueous solutions containing clay vesicles to determine if changes in nanoplate charge would destabilize the vesicles. Clay vesicles were placed in a home-made PDMS chamber. Sequentially, hydrochloric acid or sodium hydroxide was added at suitable concentrations to modify the pH of the solution from 1 to 9. After an incubation time of 1 hour at each pH value, we viewed the vesicles under a microscope. The clay vesicles remained unchanged and do not disintegrate. We also exposed the vesicles to salts of various valencies and concentrations, such as NaCl, $CaCl_2$, $MgCl_2$. The vesicles remain unchanged from zero to 1 M of salt. These observations taken together prove that electrostatic interactions play at most a minor role in keeping the montmorillonite vesicles together, and demonstrate that the particles are irreversibly aggregated in their primary van der Waals minimum.

**Additional analysis of liquid spreading on clay tablets**

Some of the liquids which were found to be completely wetting through contact angle measurements and through calculations of $S$ from surface energy components, while still spreading to a larger extent than partially wetting liquids, seem to exhibit smaller spread areas than naive expectation. Figure 10 shows an overlay of the vapor pressure (obtained from tabulated values) of the respective liquids in the right y-axis (note the log scale) of the fractional spread area plot that is reported in Fig. 7*G*. Acetone, acetaldehyde, tetrahydrofuran and methanol, the completely wetting liquids with the lowest measured fractional area spread have the highest vapor pressures among all the liquids tested. Thus the apparent low spread areas of these liquids are rationalized by larger evaporative losses during the spreading process. Note that this observation does not affect our conclusion on wetting behavior since correction for the evaporation would lead to a larger spread area for these liquids. Furthermore, there is no systematic correlation between the vapor pressure and the spread area for the other liquids. Especially note that formamide, glycerol and dimethylsulfoxide have extremely low vapor pressures. Dimethylsulfoxide spreads extensively on the surface of the clay tablet, while formamide and glycerol do not. This spreading behavior is expected since formamide and glycerol are partially wetting, while dimethylsulfoxide is completely wetting. Furthermore, water and pyridine have very similar vapor pressures, and pyridine which was found to be completely wetting, spreads more extensively on the surface of the clay tablet than water.

**Experiments with surfactants**

Water-soluble surfactants, Triton X-100 (SigmaUltra grade), Tween 20 (SigmaUltra grade), myristic acid (Sigma grade, purity ≥ 99 %), sodium octyl sulfate (purity ≥ 97%), sodium dodedecyl sulfate (purity ≥ 98%), octyltrimethylammonium bromide (purity ≥ 98%), and sodium oleate (purity ≥ 99.0 %) all purchased from Sigma-Aldrich, above their respective critical micelle concentration (CMC) were added to a suspension of clay armored bubbles in water. After an incubation time of 2 hours, we ascertained the formation of clay vesicles through optical visualization. Vesicles formed for all the surfactants tested except sodium oleate, which is only sparingly soluble in water.

**Acknowledgements.** We thank V.N. Manoharan, M.J. Russell, J.T. Trevors, J. Szostak, and D.A. Weitz for their helpful comments on the manuscript, E. Hodges for help with SEM sample preparation, and K. Ladavac for discussions. We also thank the Harvard MRSEC (DMR-0820484) and the Harvard Center for Brain Science Imaging Facility.

**Figures**

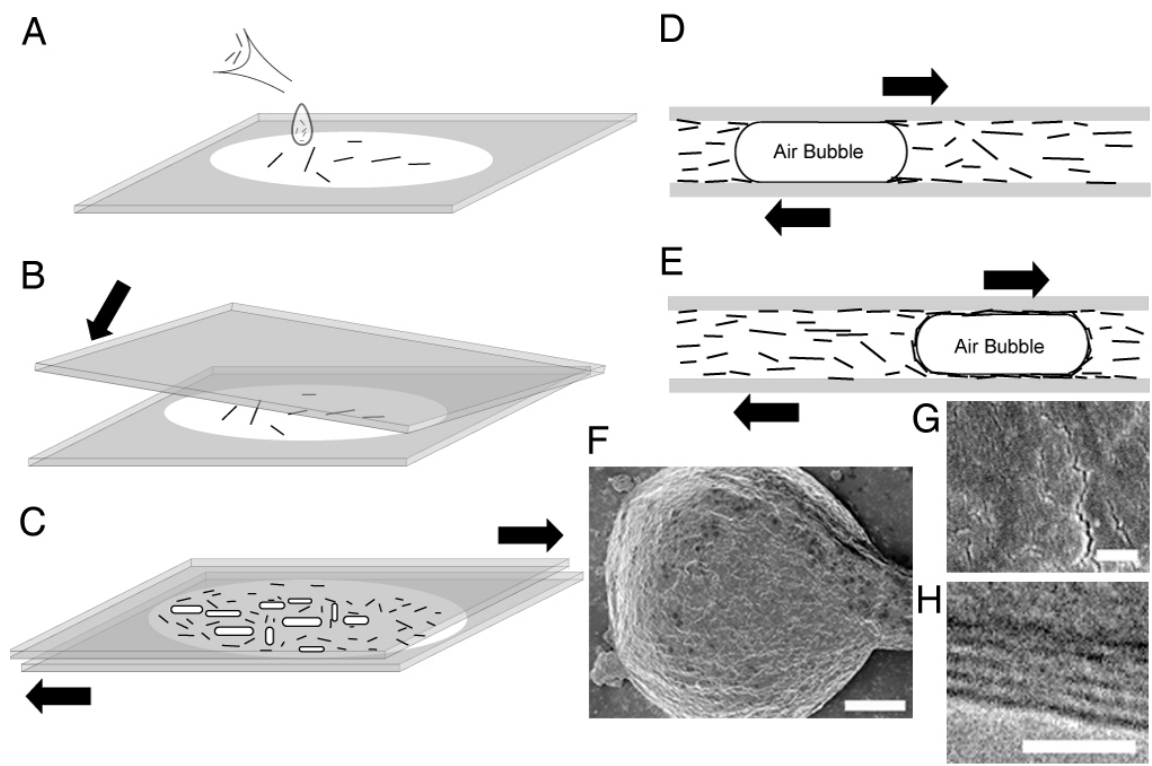

**Fig. 1 Formation of clay armored bubbles.** (A) An aqueous suspension of clay is placed on a glass slide. (B) A second glass slide is engaged on the first glass slide, trapping pockets of gas in the process. (C) The slides are pressed together manually and sheared relative to each other. Conceptualization: (D) The air bubble is deformed in the narrow gap, the clay nanoplates align in the direction of shear and some are trapped at the wedges between the air-water interface and the glass slide. (E) Eventually the bubble picks up enough particles to be fully armored. (F) SEM images of a dried clay armored bubble. (G) Higher magnification view of the clay layer, which appears smooth and continuous with no pores. (H) HRTEM image of the clay armored bubble wall showing the multilayer structure. Scale bars (F) 20 μm (G) 200 nm (H) 10 nm.

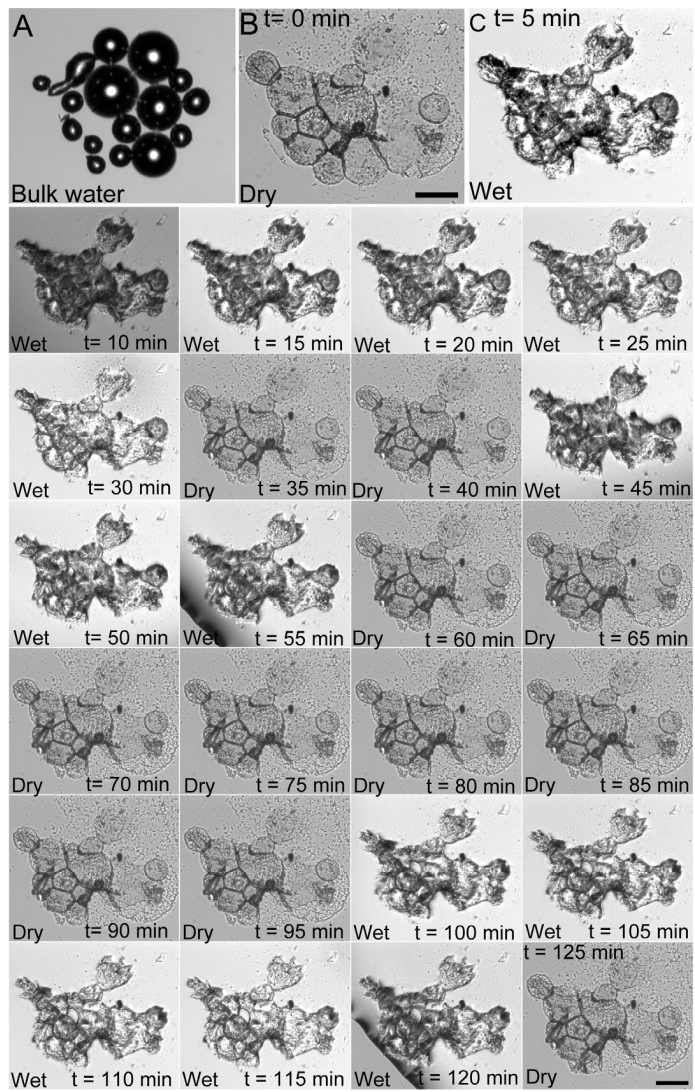

**Fig. 2 Clay armored bubbles are resilient to drying.** (A) Optical micrograph of clay armored bubbles just after preparation, note the nonspherical shapes and the contrast due to the index of refraction mismatch between air and water. (B) When water is allowed to evaporate to dryness, the clay layer on the bubbles are deposited as spheroidal aggregates on the glass slide. (C) When a fresh drop of water is placed on the dried aggregates, air is once again trapped in the interior (apparent by the reemergence of contrast). Water is allowed to evaporate to dryness before fresh drops of water are placed again at t=45 and 100 minutes. The clay aggregates' capacity to stably trap air is retained throughout the 2 hour experiment. Scale bars 45 μm.

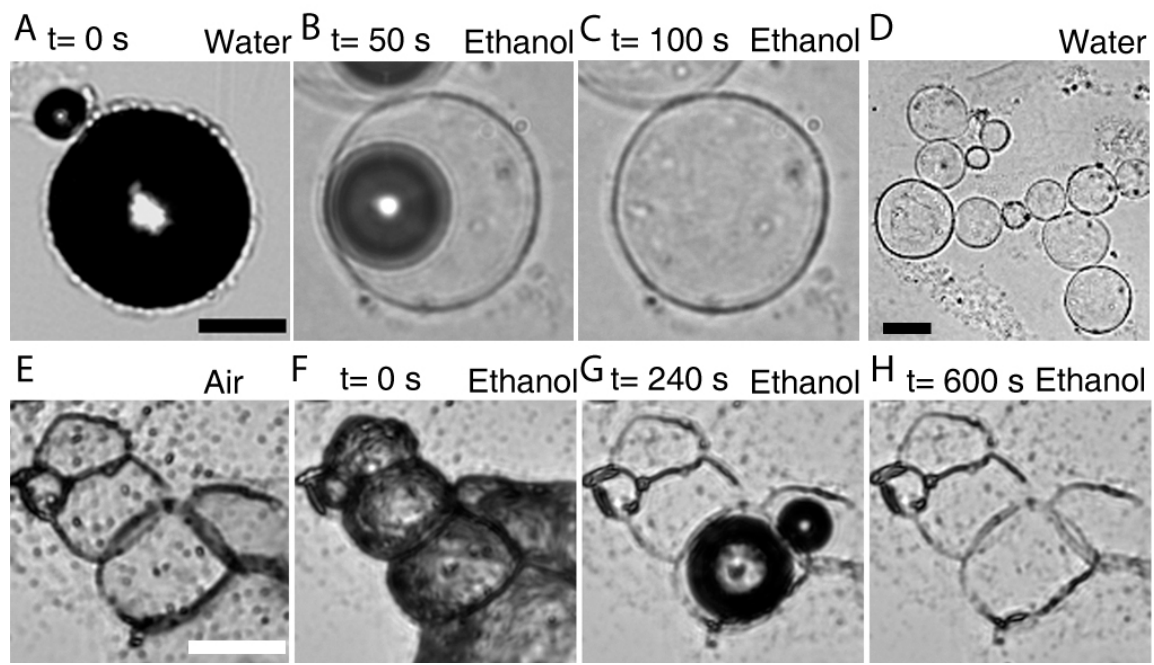

**Fig. 3 Formation of clay vesicles.** The labels identify the respective outer phases. (A) Clay armored bubbles in water. The clay layer can be discerned against the inner gas phase which appears dark. (B, C) The air bubble dissolves when the water is replaced with ethanol, producing a free-floating clay vesicle. The clay vesicle is the gray circular outline in the image. (D) A collection of clay vesicles in water. Alternate pathway: when ethanol is added to (E) dried clay armored bubbles, the (F) initially trapped gas bubbles (G) dissolve to produce (H) clay vesicles adhered to the substrate. Scale bars (A) 15 μm (D) 30 μm (E) 45 μm.

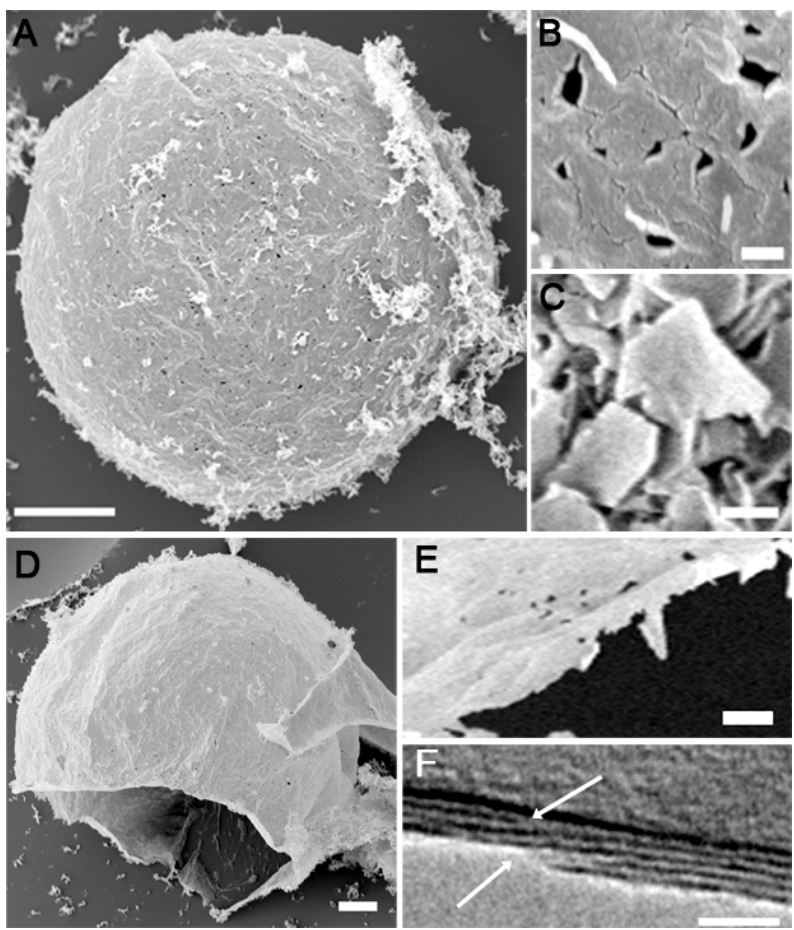

**Fig. 4 Structure of the clay vesicles**. SEM images: (A) An intact clay vesicle showing the microporous wall. Non-specifically aggregated clay particles can be seen as fluffy masses adsorbed on the upper right side of the vesicle. (B) Higher magnification view of the irregularly shaped pores on the vesicle surface. (C) Plate-like structure of free clay particles. (D) A torn clay vesicle. Non-specifically aggregated clay can be seen in the lower right hand corner. (E) A close-up of the vesicle wall. (F) HRTEM image of the vesicle edge. Individual clay nanoplates are visible as dark lines. Two apparent overlap sites are indicated with arrows. Scale bars: (A, D) 10 μm (B) 200 nm (C, E) 500 nm (F) 10 nm.

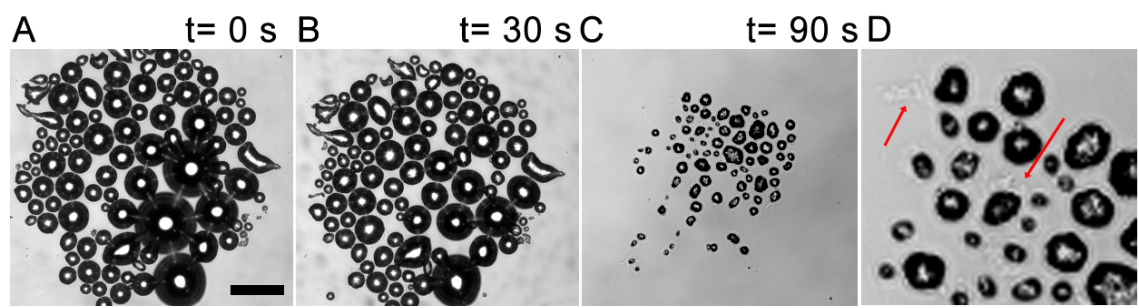

**Fig. 5 Changes in gas solubility does not explain the formation of clay vesicles.** Concentration of dissolved air in the water was reduced by degassing the water in a vacuum chamber. Gas transfer into the undersaturated water phase becomes thermodynamically favorable and thus the clay armored bubbles dissolve. Clay vesicles are not produced despite the rapid dissolution of the air bubbles. Instead as shown here in (D) which is a zoom of a section of the image in (C) aggregates of clay are ejected from bubble surfaces (red arrows), and small compact aggregates of clay are left behind after all the air has dissolved. Note that surface tension, and hence wetting properties were not changed in this experiment. (A-C) Scale bar 200 µm.

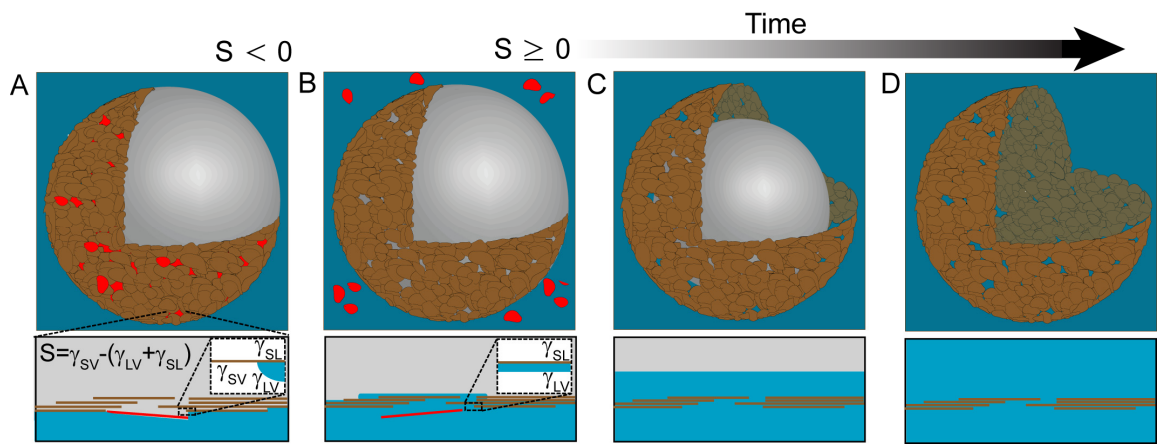

**Fig. 6 Schematic (not to scale) of the proposed mechanism of formation.** A quadrant of the shell is removed to show the air/water interface. (A) Clay plates form overlapping and aggregated shell on the air bubble surface in water. Some particles however are outside of their minimum interaction potential, and thus are only loosely aggregated (red plates). Water forms a finite contact angle on the clay nanoplates, i.e. the spreading parameter, $S$, is negative, thus the shell serves as a barrier to infiltration and the air bubble remains stable. Images on the lower row show a magnified view of a small section normal to the clay layer, emphasizing the wetting configuration, while the inset show the surface energy components for wetting (B) Liquids that completely wet clay, $S \geq 0$, form a wetting film. Particles that are in their van der Waals minimum are permanently aggregated and hence are insensitive to wetting changes. Loosely aggregated clay nanoplates however are displaced from the aggregate during the wetting transition, and so pores are formed. The pores further reduce the barrier to fluid infiltration and the liquid preferentially wets the inner surface of the clay shell, displacing the air phase. (C) Once free of the particle shell, the air bubble is no longer stable to dissolution and dissolves into the surrounding liquid. (D) All the gas is dissolved and the permanently aggregated clay shell is now a stable porous clay vesicle. The vesicle can be transferred into miscible liquids, even into partially wetting liquids which do not form vesicles, since at this stage no gas/liquid interface is present. Compare the schematic to our experimental images shown in Fig. 3A-C.

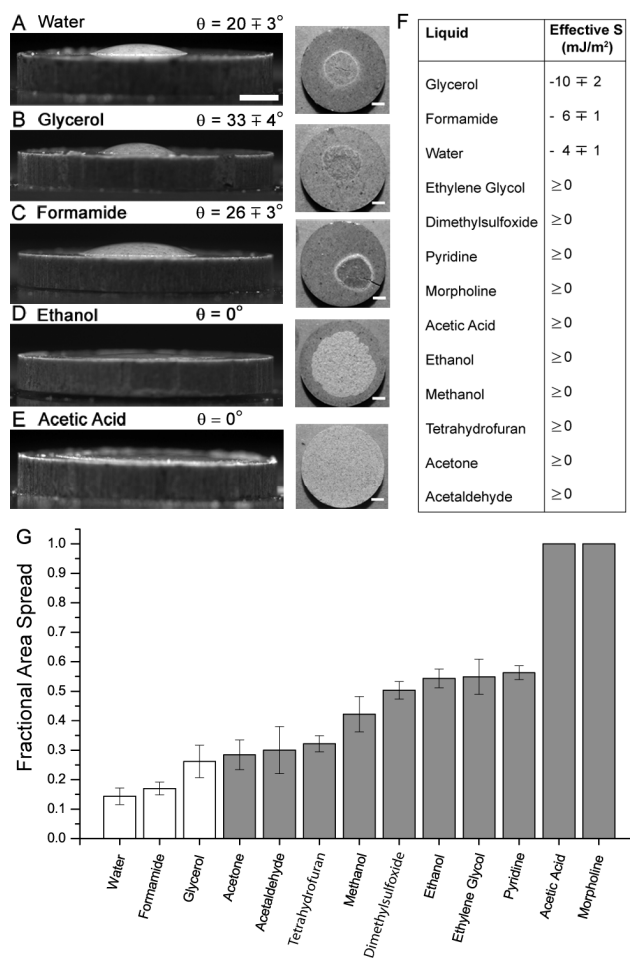

**Fig. 7 Wetting experiments on compressed montmorillonite tablets.** Left column shows representative images of the respective liquids 750 milliseconds after deposition on the clay tablets. Right column shows representative top views of the clay tablet surface 24 hours later. At this stage all liquids have either evaporated and/or imbibed into the clay tablet. Distinct intensity changes are apparent on the surface of the tablets that were in contact with liquid. Contact angles could be measured for (A) water, (B) glycerol and (C) formamide. All other liquids tested spread with very small dynamic angles, below the resolution of the experiment (<3°), as shown here for (D) ethanol and (E) acetic acid. (F) Effective spreading parameter, $S$, calculated using the Young-Dupre equation. Water, glycerol and formamide have $S < 0$, and thus are considered partially wetting on montmorillonite while all the other liquids have $S \geq 0$ and are considered

fully wetting. Wetting behavior was further quantified by measuring the fractional area spread by the liquids (area spread divided by the total area of the tablet), which is plotted in (G). Liquids that are partially wetting spread less on the surface of the tablet while liquids that are completely wetting spread more. Note that liquids that do not form vesicles (white bars) spread the least on the clay tablets. The wetting behavior of the liquids agrees with our predictions based on the hypothesized mechanism of clay vesicle formation. (A-E) Scale bars: left column 2 mm, right column 2 mm.

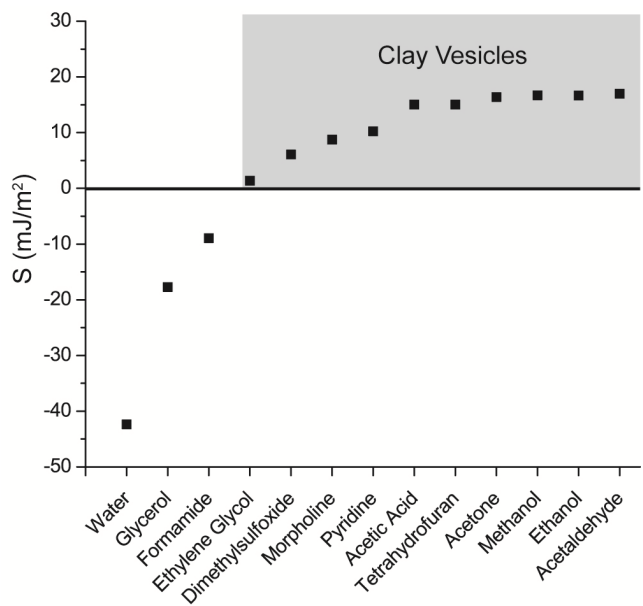

**Fig. 8 Plot of spreading parameter, *S* calculated from surface energy components.** The grey box indicates the liquids that are observed to produce clay vesicles. Consistent with our hypothesis, liquids which produce clay vesicles have $S \geq 0$.

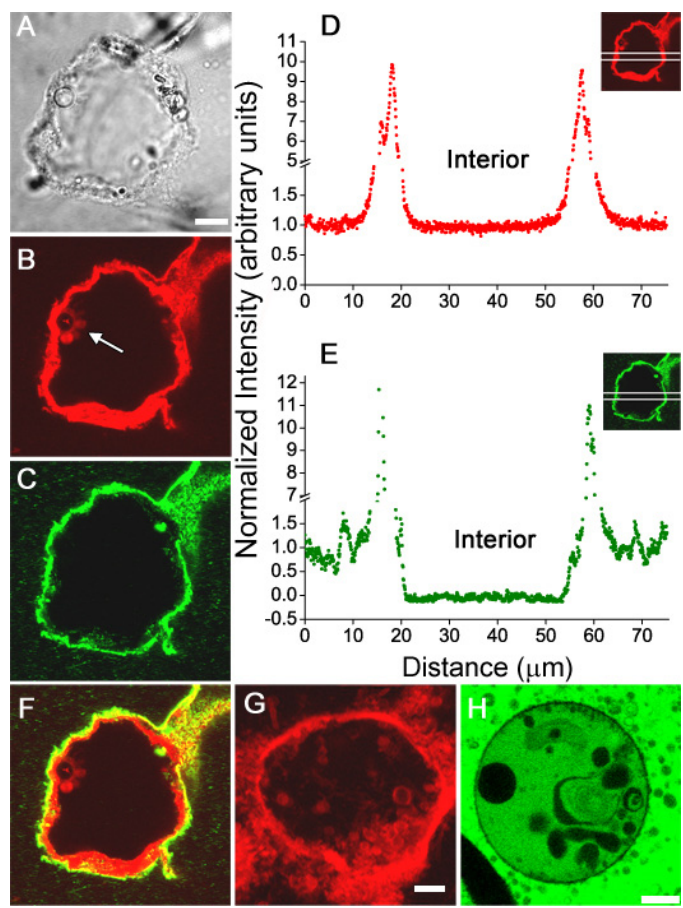

**Fig. 9 Size-selective permeability and compartmentalization of self-assembling molecules.**
(A) Transmitted light image of a surface-adhered clay vesicle. (B,C,F-H) Confocal fluorescence images: (B) Red channel: Micrometer-sized oleate liposomes, indicated by the arrow, are enclosed in the interior of the vesicle. (C) Green channel: The interior of the vesicle appears dark against a brighter background of fluorescent particles. (D, E) Normalized strip intensity profiles (insets show the strips chosen), where, (D) demonstrates that the dextran concentration is similar in the interior and exterior of the vesicle, while, (E) shows that the 110 nm diameter fluorescent particles are excluded from the interior of the clay vesicle. (F) Superimposed image of the red and green channels. Since the enclosed oleate liposomes are much larger than 110 nm, we conclude that oleate molecules enter the vesicles as monomers or small micelles which then self-assemble into liposomes. (G) Segregated liposomal populations emerge when the oleate concentration is high. Large liposomes formed in the interior of the vesicles are prevented from escaping, while those formed on the outside are barred from entering. (H) Complex internal structures can evolve, seen here in a free-floating clay vesicle. The solution is labeled with fluorescein-dextran. All scale bars 10 μm.

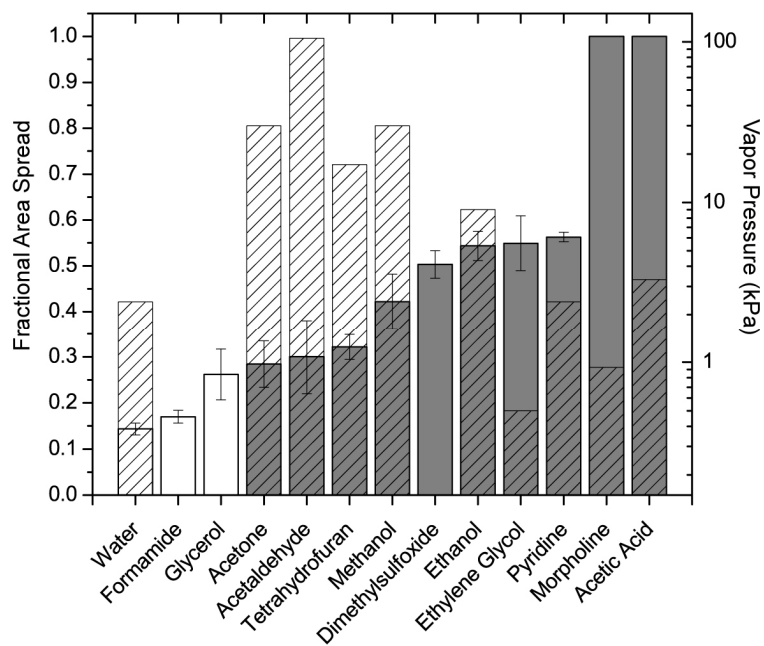

**Fig. 10** The solid bars show the fractional area spread by the liquids (x-axis) on the surface of the clay tablets (ascending order of area spread, same as Figure 7G). White bars indicate partially wetting liquids that do not form vesicles (water, formamide, glycerol). Gray bars indicate completely wetting liquids that form vesicles. The shaded bars show the vapor pressure (note the log scale on the right y-axis) of the respective liquids. The four completely wetting liquids that have the lowest measured spread areas also have the highest vapor pressures.

**Table 1:** Measured surface energies of the test liquids.

| Liquids | Surface Energy, $\gamma_{LV}$ (mJ/m$^2$) |
|---|---|
| Water | 71.9±0.2 |
| Glycerol | 62.5±0.2 |
| Formamide | 57.0±0.2 |
| Ethylene glycol | 48.0±0.2 |
| Dimethylsulfoxide | 42.5±0.2 |
| Morpholine | 38.8±0.2 |
| Pyridine | 36.5±0.2 |
| Tetrahydrofuran | 26.9±0.2 |
| Acetic Acid | 26.8±0.2 |
| Acetone | 22.8±0.2 |
| Methanol | 21.6±0.2 |
| Ethanol | 21.7±0.2 |
| Acetaldehyde | 20.4±0.2 |